%% file: main.tex
\documentclass[%
 aip,jcp,
 amsmath,amssymb,
 reprint,
]{revtex4-2}

\usepackage[a-1b]{pdfx} 

\usepackage{graphicx}
\usepackage{dcolumn}
\usepackage{bm}

\usepackage{mathptmx}
\usepackage{listings}
\usepackage{rotating} 

\usepackage{color}
\usepackage{dcolumn} 
\usepackage{bm} 
\usepackage{graphicx}
\usepackage{multirow} 
\usepackage{pifont} 
\usepackage{epsfig}
\usepackage{amsmath} 
\usepackage{subfigure}
\usepackage{float}
\usepackage{booktabs}
\usepackage{tabularx}
\usepackage{natbib}
\usepackage{gensymb}
\usepackage{rotating}
\usepackage{threeparttable}
\usepackage{comment}
\usepackage[normalem]{ulem}

\usepackage[capitalize]{cleveref}
\crefname{figure}{Figure}{Figures}
\Crefname{figure}{Figure}{Figures}
\crefname{table}{Table}{Tables}
\Crefname{table}{Table}{Tables}
\crefname{equation}{Eq.}{Eqs.}
\Crefname{equation}{Equation}{Equations}
\crefname{section}{Sec.}{Secs.}
\Crefname{section}{Section}{Sections}

\begin{document}

\author{Samragni Banerjee}
\thanks{These authors contributed equally to this work.}
\affiliation{Department of Chemistry, University of Washington, Seattle, WA 98195, USA}

\author{Run R. Li}
\thanks{These authors contributed equally to this work.}
\affiliation{
             Department of Chemistry and Biochemistry,
             Florida State University,
             Tallahassee, FL 32306-4390, USA}          

\author{Brandon C. Cooper}
\affiliation{
             Department of Chemistry and Biochemistry,
             Florida State University,
             Tallahassee, FL 32306-4390, USA}          

\author{Tianyuan Zhang}
\affiliation{Department of Chemistry, University of Washington, Seattle, WA 98195, USA}

\author{Edward F. Valeev}
\email{efv@vt.edu}
\affiliation{Department of Chemistry, Virginia Tech, Blacksburg, Virginia 24061, USA}

\author{Xiaosong Li}
\email{xsli@uw.edu}
\affiliation{Department of Chemistry, University of Washington, Seattle, WA 98195, USA}
             
\author{A. Eugene DePrince III}
\email{adeprince@fsu.edu}
\affiliation{
             Department of Chemistry and Biochemistry,
             Florida State University,
             Tallahassee, FL 32306-4390, USA}

\title{Relativistic Core-Valence-Separated Molecular Mean-Field Exact-Two-Component Equation-of-Motion Coupled Cluster Theory: Applications to L-edge X-ray Absorption Spectroscopy}

\begin{abstract}
L-edge X-ray absorption spectra for first-row transition metal complexes are obtained from relativistic equation-of-motion singles and doubles coupled-cluster (EOM-CCSD) calculations that make use of the core-valence separation (CVS) scheme, with scalar and spin–orbit relativistic effects modeled within the molecular mean-field exact two-component (X2C) framework. By incorporating relativistic effects variationally at the Dirac--Coulomb--Breit (DCB) reference level, this method delivers accurate predictions of L-edge features, including energy shifts, intensity ratios, and fine-structure splittings, across a range of molecular systems. Benchmarking against perturbative spin--orbit treatments and relativistic TDDFT highlights the superior performance and robustness of the CVS-DCB-X2C-EOM-CCSD approach, including the reliability of basis set recontraction schemes. While limitations remain in describing high-density spectral regions, our results establish CVS-DCB-X2C-EOM-CCSD as a powerful and broadly applicable tool for relativistic core-excitation spectroscopy.
\end{abstract}

\maketitle

\section{Introduction}
\input{introduction}

\section{Molecular Mean-Field Exact-Two-Component Relativistic Equation-of-Motion Coupled-Cluster with Core Valence Separation}
\input{theory}

\section{Computational Details}
\input{computational_details}

\section{Results and Discussion}
\input{results}

\section{Conclusions}
\input{conclusions}

\vspace{0.5cm}

{\bf Supporting Information} Molecular geometries, frozen orbital information, uniform shifts of simulated spectra, and calculated excitation energies and oscillator strengths at the CVS-X2C-EOM-CCSD level.

\vspace{0.5cm}

\begin{acknowledgements}
This material is based upon work supported by the U.S. Department of Energy, Office of Science, Office of Advanced Scientific Computing Research and Office of Basic Energy Sciences, Scientific Discovery through the Advanced Computing (SciDAC) program under Award No. DE-SC0022263. This project used resources of the National Energy Research Scientific Computing Center, a DOE Office of Science User Facility supported by the Office of Science of the U.S. Department of Energy under Contract No. DE-AC02-05CH11231 using NERSC award ERCAP-0024336. The development of the Chronus Quantum computational software is supported by the Office of Advanced Cyberinfrastructure, National Science Foundation (Grants No. OAC-2103717 to XL, OAC-2103705 to AED, and OAC-2103738 to EFV).
\end{acknowledgements}

\vspace{0.5cm}

\noindent {\bf DATA AVAILABILITY}\\

The data that supports the findings of this study are available within the article and its supplementary material. 

\bibliography{Journal_Long_Name,Li_Group_References,UWThesis,x2c_ref,RelCC,refs,xray_refs,Coupled_cluster_benchmark,deprince,valeev}

\end{document}

%% file: introduction.tex
\label{SEC:INTRODUCTION}

X-ray absorption spectroscopy (XAS) has emerged as an essential technique for investigating the electronic and geometric structures of molecules and materials.\cite{Stohr03_book,Mobilio14_3,Chergui14_44,Masciovecchio23_578,Yano17_book} Its ability to probe core orbitals with high precision makes XAS particularly useful for studying various chemical properties with element specificity and sensitivity to oxidation states and local coordination environments.\cite{deGroot01_1779,Li24_1118} XAS can provide detailed insights into molecular orbitals, charge transfer dynamics, and hybridization effects, establishing the approach as an indispensable tool in catalysis, materials science, and biological chemistry.\cite{Jeroen10_4754,Tada17_book,DeBeer22_5864,Yachandra09_241} 

Advances in XAS experiments, in terms of both energy and time resolution, are driving the development of advanced theoretical approaches for data interpretation.\cite{Li15_2994,Li15_4146,Dreuw18_7208,Penfold21_4276,Besley21_1527,Li20_011304} Because relativistic effects play a critical role in shaping the electronic structure of core orbitals, a sophisticated treatment of such effects is an essential component of accurate XAS simulations. Relativistic effects manifest in XAS spectra as both spectrum-wide shifts arising from scalar effects and fine-structure splitting due to spin--orbit coupling (SOC). Scalar relativistic effects can often be approximated by applying a uniform energy shift to spectra generated via non-relativistic calculations, meaning that such calculations can often provide qualitatively accurate K-edge spectra (which involve excitations from the $1s$ core level).\cite{Li11_3540,Li15_2994,Li15_4146,Coriani19_241} On the other hand, even a qualitative description of L-edge features, specifically, the L$_2$/L$_3$ edges corresponding to excitations from the $2p_{1/2}$ and $2p_{3/2}$ orbitals, which are degenerate in a non-relativistic framework,\cite{Stohr03_book,Kotani08_book} requires explicit treatment of SOC. The ability to accurately model L-edge XAS is critical, as the spectra have several desirable properties compared to K-edge XAS ({\em e.g.},  higher intensities, narrower linewidths, and longer core-hole lifetimes), which render the former approach a particularly powerful and often preferred tool for probing detailed electronic structure.

To date, a wide range of electronic structure methods, spanning density functional theory (DFT) and post-Hartree--Fock correlated approaches, have been developed and applied for modeling XAS, with several reviews providing detailed overviews of these quantum chemical methodologies.\cite{Dreuw18_7208,Kuhn20_e1433,Li20_011304,Besley21_1527,Penfold21_4276}
While DFT offers the most practical and cost-effective approach, it often suffers from large energy shifts, primarily due to self-interaction errors,\cite{Nakai07_23} and its accuracy can be sensitive to the choice of exchange–correlation functional.\cite{deSimone03_115,Li15_2994,Li18_1998,Li19_234103,Besley20_1306,Head-Gordon22_26170,Repisky23_1360}
Among the correlated wavefunction-based methods, the equation-of-motion coupled-cluster (EOM-CC) framework\cite{Emrich81_379,Bartlett89_57,Bartlett93_7029} stands out as a robust alternative with a systematically improvable description of electron correlation. Indeed, there is a long history of applying EOM-CC approaches\cite{Bartlett95_6735,Li15_4146,DePrince17_2951,Bartlett19_164117,Coriani19_3117,Matthews20_e1771448,Coriani20_8314,Bartlett21_094103,DePrince21_5438,Gomes21_3583,Cheng24_8373,Bartlett25_194306} (and the closely-related linear-response [LR] CC approach\cite{Koch15_181103}) to core-level spectroscopy. 

A significant practical challenge in simulating core-level spectra is that the relevant features are deeply embedded within the eigenvalue spectrum. As a result, one may need to calculate several hundred roots, which can be impractical when using standard iterative diagonalization methods, and this problem can become worse in large systems. As a result, many of the EOM-CC/LR-CC studies targeting core-level features\cite{Koch15_181103,Coriani19_3117,Bartlett19_164117,Matthews20_e1771448,Coriani20_8314,Gomes21_3583,Cheng24_8373} make use of the core-valence separation (CVS) scheme, originally proposed by Cederbaum, Domcke, and Schirmer,\cite{Schirmer80_206} which takes advantage of the highly localized character of core orbitals and the concomitant large energetic separation between the core and valence orbitals. In this approach, the excitation manifold is restricted to include only those configurations that involve transitions out of core orbitals, which results in several desirable outcomes. First, by eliminating pure valence transitions from the excitation space, the CVS scheme significantly reduces the floating-point costs associated with evaluating core-level spectra. Second, the approach automatically filters out spurious, poorly-described double valence excitations that exceed the ionization threshold and that may accidentally overlap spectrally with core-level features. This latter outcome is obviously useful from a physical standpoint, but it is also beneficial in that it may lead to improved convergence of iterative eigensolvers.

In this work, we present a fully relativistic CVS-EOM-CC with single and double excitations (CVS-EOM-CCSD) formalism implemented within the exact two-component (X2C) framework, where  both scalar and spin–orbit relativistic effects are incorporated variationally at the reference level. This approach builds on earlier work using the molecular mean-field (MMF) formalism and the Dirac--Coulomb--Breit Hamiltonian, which has been successfully applied to compute excitation energies, fine-structure splittings, and phosphorescence lifetimes,\cite{Li24_3408} as well as ionization potential (IP) values\cite{DePrince25_084110} and double IP values.\cite{DePrince25_104112}
In the present study, we assess the accuracy of the CVS-DCB-X2C-EOM-CCSD method in predicting key features of L-edge absorption spectra in first-row transition metal complexes, including peak shifts, intensity ratios, and fine-structure splittings. Additionally, we examine the influence of basis set recontraction, as employed in the MMF formalism, on the computed L-edge spectra.

%% file: theory.tex
\label{SEC:THEORY}

We begin with an overview of the MMF-X2C-EOM-CC approach, which supports different reference frameworks based on the treatment of relativistic two-electron interactions.\cite{Gomes18_174113,Li24_3408} Building on previous findings for valence excitations,\cite{Li24_3408} this study is focused on the most accurate Dirac--Coulomb--Breit X2C (DCB-X2C) version of MMF reference for core excitations. The approach starts by solving the four-component Dirac--Coulomb--Breit equation within the restricted-kinetic-balance condition\cite{Faegri07_book,Reiher15_book,Liu17_handbook}: 
\begin{align}
\label{EQN:DIRAC_EQUATION}
     &\begin{pmatrix}
         { \bf F}_{LL} & { \bf F}_{LS} \\
         { \bf F}_{SL} & { \bf F}_{SS}
     \end{pmatrix}
     \begin{pmatrix}
         {\bf C}^+_L &{\bf C}^-_L  \\
         {\bf C}^+_S &{\bf C}^-_S
     \end{pmatrix}\notag\\
     &=
     \begin{pmatrix}
         {\bf S} & {\bf 0}_2 \\
         {\bf 0}_2 & \frac{1}{2c^2}{\bf T}
     \end{pmatrix}
     \begin{pmatrix}
         {\bf C}^+_L &{\bf C}^-_L  \\
         {\bf C}^+_S &{\bf C}^-_S
     \end{pmatrix}
     \begin{pmatrix}
         \boldsymbol{\epsilon}^+ &{\bf 0}_2  \\
         {\bf 0}_2 &\boldsymbol{\epsilon}^-
     \end{pmatrix}
\end{align}
where, $\mathbf{S}$ and $\mathbf{T}$ are the two-component non-relativistic overlap and kinetic energy matrices, respectively, $\mathbf{F}$ denotes the four-component Fock matrix, $c$ is the speed of light, and the subscripts $L$ and $S$ refer to the large and small components of the wavefunction, respectively. The symbols $\mathbf{C}$ and $\mathbf{\epsilon}$ refer to the spinor coefficient matrices and energies, respectively, with the superscript differentiating the positive and negative energy states. The Fock matrix includes one-electron scalar and SOC relativistic effects, as well as two-electron effects captured by the DCB operator, which itself accounts for several important relativistic two-electron interactions such as spin-own orbit, spin-other orbit, spin-spin, and orbit-orbit terms, as well as their scalar products.\cite{Li21_3388,Li22_064112,Li23_171101} This DCB operator takes the form
\begin{align}
\label{EQN:DCB_OPERATOR}
    V(r_{ij}) &= \frac{1}{r_{ij}}-\frac{1}{2} \bigg( \frac{\boldsymbol{\alpha}_i \cdot \boldsymbol{\alpha}_j}{r_{ij}} 
    +\frac{(\boldsymbol{\alpha}_i \cdot {\bf r}_{ij} )(\boldsymbol{\alpha}_j \cdot {\bf r}_{ij}) }{r^3_{ij}} \bigg)    
\end{align}
where
\begin{align}
    \boldsymbol
    {\alpha}_{i,q} &= 
    \begin{pmatrix}
    {\bf 0}_2 & \boldsymbol\sigma_q \\
    \boldsymbol\sigma_q & {\bf 0}_2
    \end{pmatrix};~~
    q = \{x, y, z\}
    \label{eq:breit}
\end{align}
Here, $\boldsymbol\sigma$ denotes the Pauli spin matrices, $\{i,j\}$ are electron labels and ${\bf 0}_2$ represents the $2\times2$ zero matrix. 

Equation \ref{EQN:DIRAC_EQUATION} contains both electronic and positronic degrees of freedom, but the latter are not directly relevant for quantum chemistry applications. For correlated many-body methods, finding a way to decouple the electronic and positronic components of the problem is desirable, as it automatically eliminates artifacts associated with the positronic states\cite{VRG:sucher:1980:PRA,VRG:sucher:1984:IJoQC} as well as increases efficiency by treating correlation in the space of positive-energy states represented by two-component states (spinors), rather than four-component states (bispinors). This is the motivation behind the X2C approach.  \cite{Dyall97_9618,Dyall98_4201,Enevoldsen99_10000,Dyall01_9136,Cremer02_259,Liu05_241102,Peng06_044102,Cheng07_104106,Saue07_064102,Peng09_031104,Liu10_1679,Liu12_154114,Reiher13_184105,Li16_3711,Li16_104107,Repisky16_5823,Li17_2591,Cheng21_e1536,Li22_2947,Li22_2983,Li22_5011,Li24_3408,Li24_7694,Li24_041404} In the MMF flavor of X2C, one begins by solving Eq.~\ref{EQN:DIRAC_EQUATION} for four-component molecular spinors. The electronic spinors define a unitary four-component-to-two-component transformation that is applied to the Fock operator and the Coulomb part Eq.~\ref{EQN:DCB_OPERATOR} for subsequent use in correlated calculations, which are carried out in the two-component basis. As such, two-electron relativistic effects enter the correlated calculation through the Fock operator ({\em i.e.}, in a mean-field way), while it is assumed that the correlation effects stemming from the Breit operator can be neglected.

The ground-state X2C-CC wave function is
\begin{align}
| \Psi_0\rangle = \exp(\hat{T})|\Phi_0 \rangle
\end{align}
where $|\Phi_0\rangle$ is a two-component reference configuration, which is a determinant of electronic molecular spinors obtained from solving Eq.~\ref{EQN:DIRAC_EQUATION}, followed by the X2C transformation. At the CC with single and double excitations level (CCSD)\cite{Bartlett82_1910}, the cluster operator, $\hat{T}$, is defined by
\begin{align}
    \hat{T} = \sum_{ia} t_i^a \hat{a}^\dagger_a \hat{a}_i + \frac{1}{4} \sum_{ijab} t_{ij}^{ab} \hat{a}^\dagger_a \hat{a}^\dagger_b \hat{a}_j \hat{a}_i 
\end{align}
where $t_i^a$ and $t_{ij}^{ab}$ are the cluster amplitudes, the symbols $\hat{a}_i$ and $\hat{a}^\dagger_a$ refer to annihilation and creation operators for molecular spinors $i$ and $a$, respectively, and the labels $i$ and $j$ versus $a$ and $b$ refer to spinors that are occupied or unoccupied in $|\Phi_0\rangle$. The cluster amplitudes are determined in the usual projective way, after which the ground-state CC energy is given by the expectation value
\begin{equation}
    \label{EQN:CC_ENERGY}
    E_{0} = \langle \Phi_0 | {\bar{H}} | \Phi_0 \rangle
\end{equation}
where $\bar{H} = \exp(-\hat{T}) \hat{H}\exp(\hat{T})$ is the similarity-transformed Hamiltonian.

Once the ground-state CC amplitudes have been determined, the energies of core excited states can be determined within the EOM-CC formalism\cite{Bartlett93_7029}, as eigenvalues of the similarity-transformed Hamiltonian. The $K$th excited state ($K > 0$) is parametrized by left- and right-hand EOM-CC functions defined by
\begin{align}
    \langle \tilde{\Psi}_K | &= \langle \Phi_0 | \hat{L}_K \exp(-\hat{T})
\end{align}
and
\begin{align}
    |\Psi_K\rangle &= \hat{R}_K \exp(\hat{T}) |\Phi_0\rangle
\end{align}
that satisfy left- and right-hand eigenvalue equations of the form
\begin{align}
\label{EQN:EOM_EIGENVALUE_LEFT}
   \langle \Phi_0 | \hat{L}_K  \bar{H}  &= E_K  \langle \Phi_0 | \hat{L}_K
\end{align}
and
\begin{align}
    \label{EQN:EOM_EIGENVALUE_RIGHT}
    \bar{H} \hat{R}_K |\Phi_0 \rangle &= E_K \hat{R}_K |\Phi_0 \rangle
\end{align}
where $E_K$ is the energy of state $K$. Excitation energies are then given by $\omega_K = E_K - E_0$.

As already mentioned, solving Eqs.~\ref{EQN:EOM_EIGENVALUE_LEFT} and \ref{EQN:EOM_EIGENVALUE_RIGHT} for core excited states can be challenging for standard iterative eigensolver algorithms because the core states lie far from the periphery of the spectrum of $\bar{H}$. As such, we invoke the CVS approximation, which imposes a special structure on $\hat{R}_K$ and $\hat{L}_K$ so that Eqs.~\ref{EQN:EOM_EIGENVALUE_LEFT} and \ref{EQN:EOM_EIGENVALUE_RIGHT} can be solved for the core states without considering any of the lower-lying valence states. Among the different variants of the CVS approximation available in the literature, this work follows the formulation in Ref. \citenum{Coriani19_3117}, where the occupied molecular spinors are partitioned into three subspaces: (i) frozen core occupied spinors that are not correlated in the CC/EOM-CC parts of the calculations, (ii) correlated core occupied spinors that are relevant for the core excited states of interest, and (iii) valence occupied spinors. The EOM-CC excitation operators are then defined by
\begin{align}
\label{EQN:CC_LEFT_CVS}
\hat{L}_K &= l_{K,0} + \sum_{Ia} l_{K,a}^{\phantom{K,}I} \hat{a}_I^{\dagger} \hat{a}_a + \frac{1}{4} \sum_{IJab} l_{K,ab}^{\phantom{K,}IJ} \hat{a}_I^{\dagger} \hat{a}_J^{\dagger} \hat{a}_b \hat{a}_a \nonumber \\
&+ \frac{1}{2} \sum_{Ijab} l_{K,ab}^{\phantom{K,}Ij} \hat{a}_I^{\dagger} \hat{a}_j^{\dagger} \hat{a}_b \hat{a}_a
\end{align}
and
\begin{align}
\label{EQN:CC_RIGHT_CVS}
\hat{R}_K &= r_{K,0} + \sum_{Ia} r_{K,I}^{\phantom{K,}a} \hat{a}_a^{\dagger} \hat{a}_I + \frac{1}{4} \sum_{IJab} r_{K,IJ}^{\phantom{K,}ab} \hat{a}_a^{\dagger} \hat{a}_b^{\dagger} \hat{a}_J \hat{a}_I \nonumber \\
&+ \frac{1}{2} \sum_{Ijab} r_{K,Ij}^{\phantom{K,}ab} \hat{a}_a^{\dagger} \hat{a}_b^{\dagger} \hat{a}_j \hat{a}_I
\end{align}
where $l_{K,0}$, $r_{K,0}$, etc. are expansion coefficients obtained by solving Eqs.~\ref{EQN:EOM_EIGENVALUE_LEFT} and \ref{EQN:EOM_EIGENVALUE_RIGHT}. Here, the labels $I$ and $J$ refer to the the correlated core occupied spinors, and the label $j$ refers to the correlated valence occupied spinors. Because all transition operators in Eqs.~\ref{EQN:CC_LEFT_CVS} and Eqs.~\ref{EQN:CC_RIGHT_CVS} involve at least one core orbital, configurations associated with valence-only transitions will not enter the associated eigenvalue problems.

%% file: computational_details.tex
\label{SEC:COMPUTATIONAL_DETAILS}

The CVS-DCB-X2C-EOM-CCSD method has been implemented in the ChronusQuantum software package.\cite{Li20_e1436} Tensor operations arising in the CCSD amplitude equations and the construction of EOM-CCSD sigma vectors are powered by the TiledArray tensor library\cite{TiledArray,VRG:calvin:2015:I15WIAAA}
based on the MADNESS task-parallel runtime.\cite{VRG:harrison:2016:SJSC} Energies and left- and right-hand EOM-CC vectors for core excited states were determined from a Davidson\cite{Davidson75_87,Davidson89_49} eigensolver adapted for complex arithmetic. All calculations were performed with a Davidson threshold of 10$^{-7}$ au for energy convergence and 10$^{-5}$ au for the residual norm. {All X2C-EOM-CCSD spectra were generated by applying a Lorentzian convolution, and the full width at half maximum (FWHM) are listed in the Supporting Information. Experimental and previously published TDDFT spectra were extracted using PlotDigitizer \cite{WebPlotDigitizer} and displayed for comparison. All theoretical spectra are shifted so that the highest peak in the L$_3$ edge is aligned to the experimental one. }

All calculations were carried out using the frozen core approximation, where deeply-bound core orbitals are excluded from the correlated portions of the calculations (both the CC and EOM-CC parts). Additional details regarding the specific core orbitals that were excluded from the correlation treatments are provided in the Supporting Information.

Molecular geometries for SiCl$_{4}$, TiCl$_{4}$, and CrO$_{2}$Cl$_{2}$ were obtained from Ref.~\citenum{Li18_1998}. The geometry for VOCl$_{3}$ was obtained from Ref.~\citenum{Li19_234103}. All geometric parameters are reproduced in the Supporting Information.

The CVS-DCB-X2C-EOM-CCSD implementation was validated against the CVS-EOM-CCSD data for Ar atom obtained from Ref.~\citenum{Coriani20_8314}. For this study, we used the same uncontracted 6-311(2$+$,$+$)G(p,d) basis set, supplemented with additional Rydberg-type functions, that was used in that work.
All calculations on the metal complexes were performed with the x2c-TZVPall-2c basis set.\cite{Weigend17_3696,Weigend19_16658} We employed the same basis set recontraction scheme used in our previous work on valence excitations.\cite{Li24_3408} The four-component Dirac--Hartree--Fock calculations were performed with an uncontracted x2c-TZVPall-2c basis, which was then recontracted after the X2C transformation. The subsequent CC and CVS-EOM-CC calculations were performed within the recontracted basis.

%% file: results.tex
\label{SEC:RESULTS}

\subsection{Validation}
\label{SEC:Validation}

\begin{figure}[!htpb]
\includegraphics[width=\linewidth]{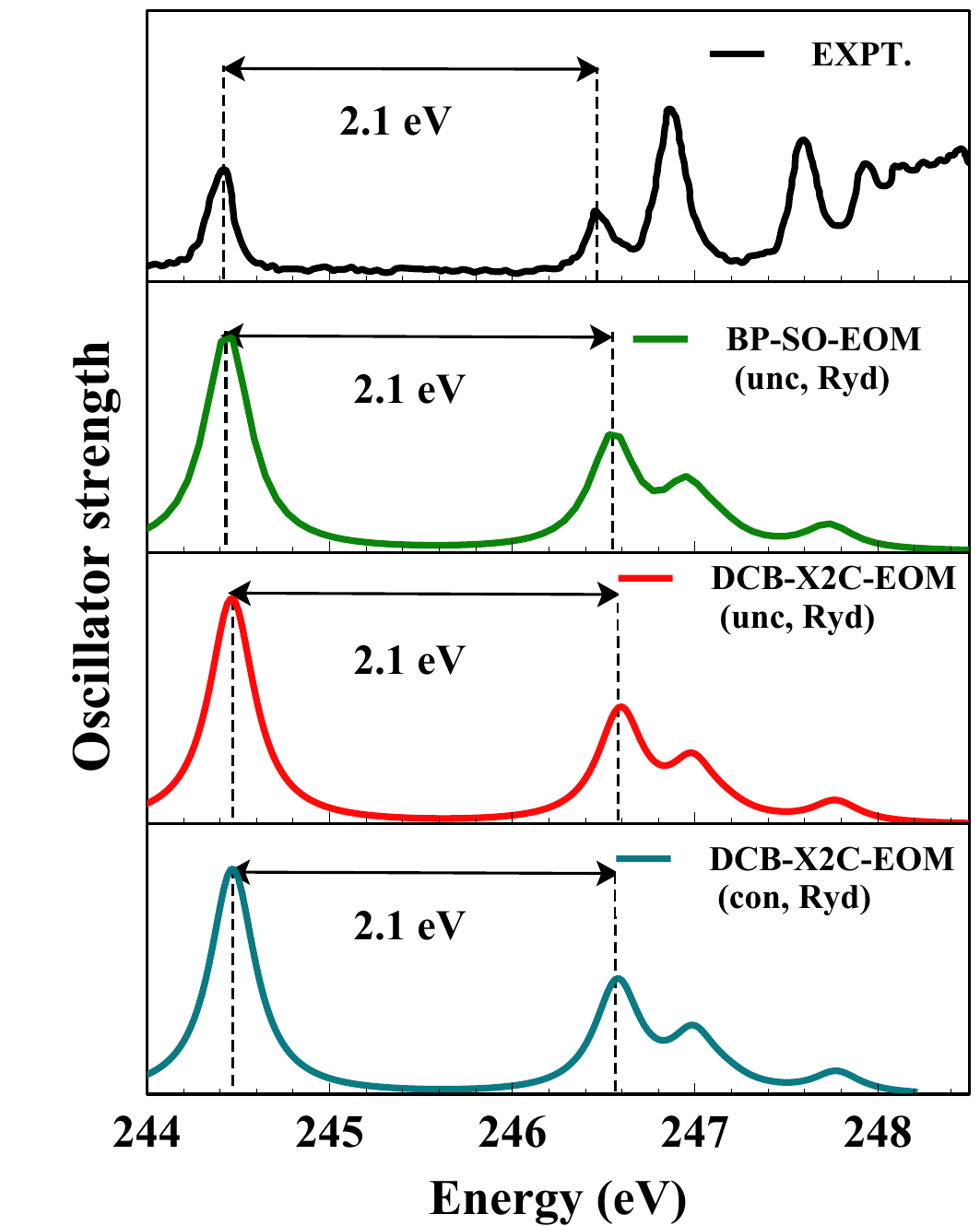}
\caption{Comparison of the experimental L$_{2,3}$-edge spectra of Argon with theoretical spectra computed using CVS-BP-SO-EOM-CC and CVS-DCB-X2C-EOM-CC methods. The second and third panels display spectra computed with the uncontracted 6-311(2+,+)G(p,d) basis set, supplemented by Rydberg-type functions as outlined in Ref. \citenum{Coriani20_8314}, while the fourth panel presents results obtained using the contracted version of the same basis. The applied energy shifts are as follows: CVS-BP-SO-EOM: $+$0.7 eV, CVS-DCB-X2C-EOM with the uncontracted basis: {$-$0.6} eV, and CVS-DCB-X2C-EOM with the contracted basis: {$-$0.9} eV. The CVS-BP-SO-EOM-CC results have been reproduced from Ref. \citenum{Coriani20_8314}. The experimental spectrum is taken from Ref.~\citenum{Oshio68_1303}.
}\label{fig:Ar_L}
\end{figure}

We begin by benchmarking our implementation of CVS-DCB-X2C-EOM-CC  against data from Ref.~\citenum{Coriani20_8314} generated with a perturbative spin--orbit CVS-EOM-CC method that utilized the Breit-Pauli (BP) Hamiltonian {(CVS-BP-SO-EOM-CC)}. 
For this purpose, we consider argon atom, which features optically allowed transitions from $2p$ to $ns$ ($n \geq 4$) or $nd$ ($n \geq 3$) orbitals. The $2p$ hole can have a total angular momentum quantum number $j = 3/2$ or $j = 1/2$, resulting in two distinct series. The first panel of \cref{fig:Ar_L} depicts the experimental spectrum,\cite{Oshio68_1303} {where the first two spectral bands, located at 244.4 eV and 246.5 eV}, correspond to the $2p_{3/2} \to 4s$ and $2p_{1/2} \to 4s$ transitions, respectively. The third band at 247 eV is attributed to transitions from $2p_{3/2} \to 5s, 3d$ orbitals, while the fourth band, around 247.5 eV corresponds to $2p_{3/2} \to 6s, 4d$ excitations. 

The second and third panels of \cref{fig:Ar_L} depict the simulated spectra generated via CVS-BP-SO-EOM-CC and CVS-DCB-X2C-EOM-CC, using the uncontracted basis, augmented by Rydberg functions, as described in Ref.~\citenum{Coriani20_8314} and in the previous section. We find good agreement between these spectra, in terms of both peak spacings and intensities. 
Both methods recover the spacing between $2p_{3/2} \to 4s$ and $2p_{1/2} \to 4s$ transitions in the experimental spectrum (2.1 eV). As for the absolute peak positions, the CVS-DCB-X2C-EOM-CC spectrum requires a {$-$0.6} eV shift to align it with the experimental one, whereas CVS-BP-SO-EOM-CC requires a slightly larger shift, of opposite sign (0.7 eV). The low intensities of the third and fourth peaks in the simulated spectra, as compared to experiment, may indicate that more diffuse basis functions may be necessary to accurately describe the transition intensities for core excited states involving higher-lying unoccupied orbitals.

The fourth panel of \cref{fig:Ar_L} provides additional CVS-DCB-X2C-EOM-CC data generated using a contracted basis set. Previous work\cite{Li24_3408} has  demonstrated that the DCB-X2C-EOM-CC method provides reliable results for valence excitations using basis sets that are recontracted after the X2C transformation. We make a similar observation here for core excitations. The spectrum obtained with the contracted basis aligns closely with that from the uncontracted basis, with a small additional shift of $-$0.3 eV (which results in an overall {$-$0.9} eV shift to align with the experimental result). Apart from this minor adjustment, the calculation carried out in the contracted basis reproduces all spectral features, including peak positions and intensities, suggesting that basis set recontraction has a minimal impact on the quality of L-edge absorption spectra. Consequently, all subsequent results discussed in the section were obtained from calculations carried out within the recontracted basis.

\subsection{SiCl$_{4}$}
\label{SEC:Si}

\begin{figure}[!htpb]
\includegraphics[width=\linewidth]{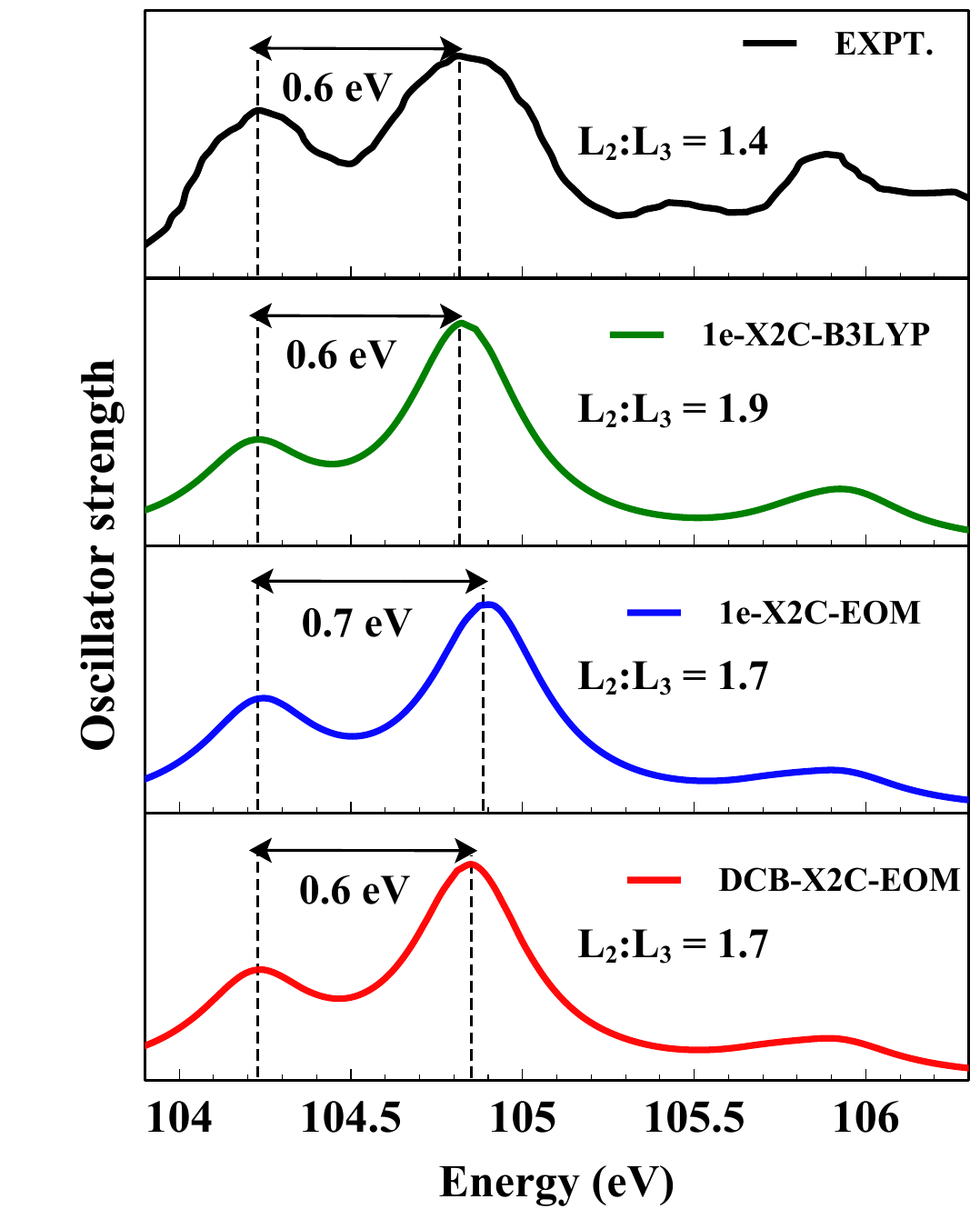}
 \caption{ Comparison of L$_{2,3}$-edge spectra of SiCl$_{4}$ computed with two variants of CVS-X2C-EOM-CC using x2c-TZVPall-2c basis. Also shown in the first panel is the experimental spectrum\cite{Tse87_33}. The 1e-X2C-TDDFT spectrum in the second panel obtained with the B3LYP functional and aug-cc-pVTZ basis has been reproduced from Ref. \citenum{Li18_1998}. The applied energy shifts are as follows: 1e-X2C-B3LYP: $+$6 eV, CVS-DCB-X2C-EOM: $-$0.6 eV, and 1e-X2C-EOM: {$-$0.7} eV. The experimental spectrum is taken from Ref.~\citenum{Tse87_33}}\label{fig:Si_L}.
\end{figure}

We now turn to the simulation of the L$_{2,3}$ edge of SiCl$_{4}$, which is presented in \cref{{fig:Si_L}}. This spectrum exhibits several characteristic features that can be attributed to the mixing of the Si $2p$ orbitals with the ligand environment.\cite{Tse87_33} The most prominent peaks appearing at 104.2 eV and 104.8 eV arise due to the excitation of electrons from Si $2p_{3/2}$ and $2p_{1/2}$ orbitals, respectively, to unoccupied molecular orbitals with Si-Cl antibonding character.\cite{Wen95_285} 
The L$_{2}$-L$_{3}$ splitting is only $\sim$0.6 eV, which makes the interpretation of the L$_{2,3}$ spectrum of Si particularly challenging due to the overlap of features corresponding to both edges.

In addition to the experimental spectrum, \cref{{fig:Si_L}} provides simulated spectra for the L$_{2,3}$ edge of SiCl$_4$ computed using X2C time-dependent DFT (X2C-TDDFT)\cite{Li11_3540,Li17_2591,Li18_169,Li18_2034,Li18_1998} and CVS-X2C-EOM-CC using two formulations of the X2C approach. For CVS-X2C-EOM-CC, we provide data generated using the MMF/DCB approach (panel three of \cref{{fig:Si_L}}), as well as data from simulations carried out using the more approximate one-electron X2C (1e-X2C) approach (panel four of \cref{{fig:Si_L}}). In the 1e-X2C framework, one-electron scalar and one-electron spin-orbit relativistic effects are accounted for via a four-component-to-two-component transformation carried out prior to the self-consistent field (SCF) step. The SCF and subsequent correlation treatments are carried out within the two-component basis.  To account for missing two-electron spin-orbit coupling effects, we employ the row-dependent DCB-parametrized screened nuclear spin--orbit (SNSO) factor.\cite{Boettger00_7809,Li21_3388,Li22_064112,Li23_171101,Li23_5785} The X2C-TDDFT spectrum presented in the second panel of \cref{{fig:Si_L}}, which was reproduced from Ref.~\citenum{Li18_1998}, was generated using this same 1e-X2C framework and the B3LYP functional. 

Comparing the simulated spectra to the experimentally obtained one, we note that 1e-X2C-B3LYP and CVS-DCB-X2C-EOM-CC recover the 0.6 eV splitting between the L$_{2}$-L$_{3}$ features, whereas CVS-1e-X2C-EOM-CC predicts a slightly larger splitting (0.7 eV). In terms of absolute peak positions,  the CVS-X2C-EOM-CC spectra require a $-$0.6 eV shift to align them with the experimental spectrum, regardless of the relativistic treatment, whereas 1e-X2C-B3LYP requires an order-of-magnitude-larger shift of 6 eV. Experimentally, the relative intensity of the L$_2$/L$_3$ edge features at 104.8 eV and 104.2 eV is 1.4. 1e-X2C-B3LYP predicts a much higher ratio of 1.9, whereas both CVS-X2C-EOM-CC approaches predict a slightly improved ratio of 1.7. Overall, both CVS-X2C-EOM-CC methods produce nearly identical spectra, with the X2C-DCB framework providing slightly more accurate results. As such, the remaining CVS-X2C-EOM-CC calculations discussed below consider only the X2C-DCB framework. 

\subsection{TiCl$_{4}$}
\label{SEC:Ti}

\begin{figure}[!htpb]
\includegraphics[width=\linewidth]{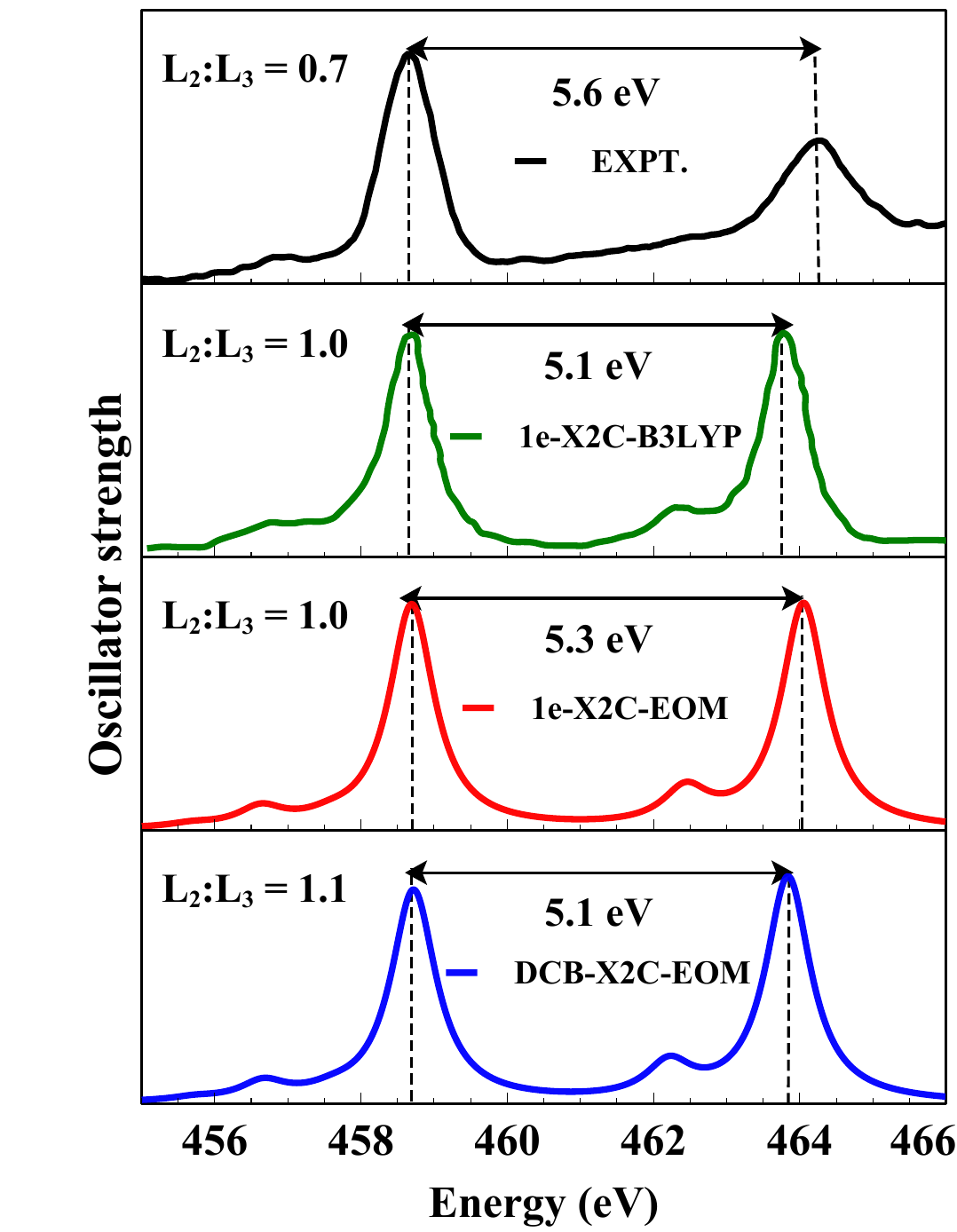}
 \caption{Comparison of the L$_{2,3}$-edge spectra of TiCl$_{4}$ computed using 1e-X2C-TDDFT and CVS-DCB-X2C-EOM-CC with the x2c-TZVPall-2c basis set, alongside the experimental spectrum in the uppermost panel (reproduced from Ref.~\citenum{Hitchcock93_1632}). The 1e-X2C-TDDFT spectrum in the second panel obtained with the B3LYP functional and aug-cc-pVTZ basis has been reproduced from Ref. \citenum{Li18_1998}. The applied energy shifts are as follows: 1e-X2C-B3LYP: $+$9.5 eV, {CVS-1e-X2C-EOM-CC:$-$2.9 eV, and CVS-DCB-X2C-EOM-CC: $-$2.7 eV.}
 }\label{fig:Ti_L}
\end{figure}

Although TiCl$_{4}$ has a tetrahedral symmetry similar to SiCl$_{4}$, it features a simpler spectrum due to the relatively localized nature of the Ti $2p$ orbitals. As shown in \Cref{{fig:Ti_L}}, two prominent features appear around 459 eV and 464 eV, corresponding to the L$_{3}$ and L$_{2}$ peaks, respectively, and arise due to $2p \rightarrow 3d$ transitions in Ti. Additionally, both edges display weak shoulder features at 457 eV and 462 eV, that can be attributed to crystal field splitting of the unoccupied $3d$ orbitals.

The simulated spectra show reasonable qualitative agreement with the experimentally obtained spectrum. Unlike the case of SiCl$_4$, though, we observe significant quantitative differences between simulation and experiment. First, {spectra derived from 1e-X2C-TDDFT, require much more substantial energy shifts (9.5 eV) than those from CVS-DCB-X2C-EOM-CC and CVS-1e-X2C-EOM-CC calculations ($-$2.7 eV and $-$2.9 eV, respectively).} Second, X2C-TDDFT and CVS-DCB-X2C-EOM-CC both underestimate the L$_2$--L$_3$ splitting from experiment (5.6 eV) {by 0.5 eV, whereas CVS-1e-X2C-EOM-CC underestimates this value by 0.3 eV.} Similar to the case of SiCl$_4$, we observe substantial discrepancies in experimental and simulated relative intensities of the L$_{2,3}$ edge features. The experimentally obtained L${_2}$/L${_3}$ ratio is 0.7, whereas 1e-X2C-B3LYP, CVS-1e-X2C-EOM-CC, and CVS-DCB-X2C-EOM-CC predict higher ratios of 1.0, 1.0, and 1.1, respectively.

    The similar performance of CVS-1e-X2C-EOM-CC and CVS-DCB-X2C-EOM-CC suggests that the row-dependent DCB-parametrized SNSO scaling factor\cite{Li23_5785} in the 1e-X2C scheme does a good job of mimicking two-electron spin-orbit coupling effects for this system. We can further quantify the importance of such terms by comparing L$_{2,3}$-edge spectra computed with different relativistic Hamiltonians within the MMF-X2C scheme. 
Table S8 in the Supporting Information tabulates the peak shifts, L$_2$--L$_3$ splitting values, and L${_2}$/L${_3}$ peak intensity ratios computed using different relativistic treatments (1e-X2C, DC-X2C, and DCB-X2C), as well as with different numbers of frozen occupied orbitals. For the frozen orbital scheme used to generate the data in Fig.~\ref{fig:Ti_L} (Ti $1s$, $2s$ and Cl $1s$, $2s$, $2p$ frozen), the energy shifts required to align the calculated and experimental spectra differ by less than 0.3 eV accross relativistic treatments, as do the L$_2$--L$_3$ splitting values themselves. The L${_2}$/L${_3}$ peak intensity ratios are also similar for each methods. Regarding the correlation space, we find that the spectra in Fig.~\ref{fig:Ti_L} are reasonably well converged with respect to the number of frozen occupied orbitals. CVS-DCB-X2C-EOM-CC the overall shift and L$_2$--L$_3$ splittings obtained with the present freezing scheme differ from those obtained when correlating all occupied orbitals differ by only 0.1 eV. On the other hand, freezing additional occupied orbitals leads to the need for significantly larger overall shifts to align the spectra with experiment. Put together, these data suggest that discrepancies with respect to experiment are likely due to the lack of high-order correlation effects ({\em i.e.}, those due to triple and higher excitations), as well as the choice of basis set. Based on these data, in the following discussion of VOCl$_3$ and CrO$_2$Cl$_2$, we consider simulations that freeze only those orbitals with energies lower than the central atom $3s$ orbitals, with the exception of the $2p$ orbitals that are responsible for the L$_{2,3}$ transitions.

\subsection{VOCl$_{3}$}
\label{SEC:V}

\begin{figure}[!htpb]
\includegraphics[width=\linewidth]{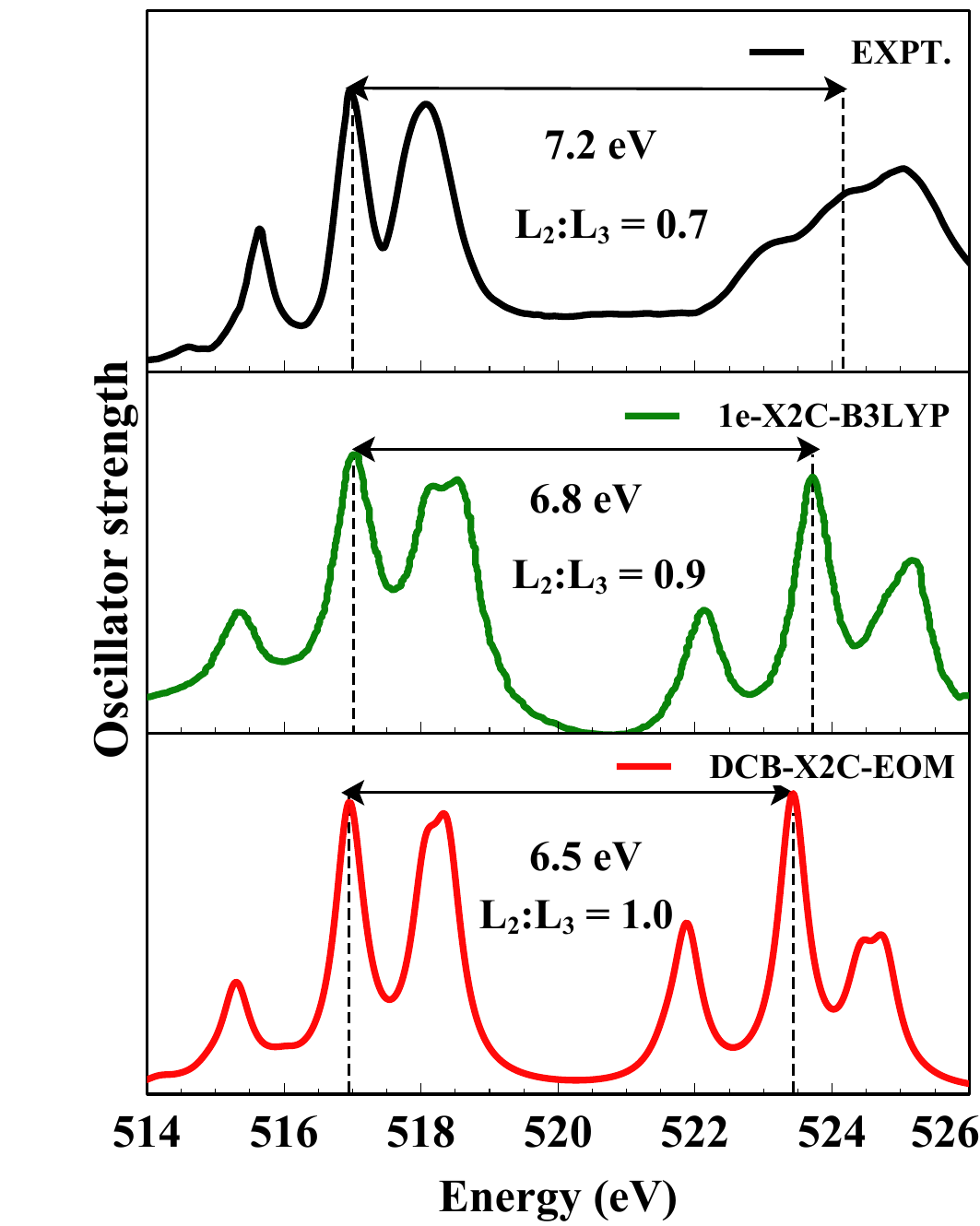}
 \caption{The computed L$_{2,3}$-edge spectra of VOCl$_{3}$ with 1e-X2C-TDDFT and DCB-X2C-EOM-CC using x2c-TZVPall-2c basis, compared to the experimental spectrum (reproduced from Ref.~\citenum{Prince09_2914}). The 1e-X2C-TDDFT spectrum in the second panel obtained with the B3LYP functional and aug-cc-pVTZ basis has been reproduced from Ref. \citenum{Li19_234103}. The applied energy shifts are as follows: 1e-X2C-B3LYP: $+$4.3 eV, and DCB-X2C-EOM: {$-$3.0} eV.
 }\label{fig:V_L}
\end{figure}

We now move on to the L$_{2,3}$-edge spectrum for VOCl$_3$ (see \Cref{{fig:V_L}}), which is more intricate than those of SiCl$_{4}$ and TiCl$_{4}$ discussed above and exhibits several characteristic features.  An accurate interpretation of this spectrum requires both spin--orbit coupling and crystal field effect considerations, as both contribute to the structure of the spectrum. Spin--orbit coupling splits the vanadium $2p$ core level into 2p$_{3/2}$ (L$_3$) and 2p$_{1/2}$ (L$_2$) components, whereas crystal field effects reduce the symmetry of the molecule from $T_d$ to $C_\text{3v}$, which splits the V $3d$ orbitals into three virtual valence levels. The interplay of these effects gives rise to a dense manifold of transitions across the L$_3$ and L$_2$ bands. We refer the readers to Ref.\citenum{Prince09_2914} for more detailed discussion of these transitions.

Comparing simulated spectra in \Cref{{fig:V_L}} to experiment, we first note that 1e-X2C-B3LYP and CVS-DCB-X2C-EOM-CC require opposite-in-sign shifts to align the most intense feature at 517 eV, where the shift required by CVS-DCB-X2C-EOM-CC ({$-$3.0 eV}) is slightly smaller in magnitude than that by 1e-X2C-B3LYP ($+$4.3 eV). With these applied shifts, the simulated L$_3$-edge spectra show good qualitative agreement with experiment, but several quantitative differences are worth discussing. From experiment, the spacings between the first and second peaks and the second and third peaks in the low-energy part of the L$_3$ region are 1.4 eV and 1.1 eV, respectively. We find that 1e-X2C-B3LYP over estimates both of these splittings ({1.7} eV and {1.3 eV}), while CVS-DCB-X2C-EOM-CC yields an improved splitting between the first two peaks and a comparable-in-quality splitting between the second and third peak ({1.6} eV and {1.3} eV, respectively). Additionally, with respect to relative peak intensities, the CVS-DCB-X2C-EOM-CC data more closely resemble experiment, as compared to the 1e-X2C-B3LYP data. {The pre-peak near 515 eV is also observed in the CVS-DCB-X2C-EOM-CC spectrum. }

The quality of simulated spectra at higher energies ($>521$ eV) is worse than that observed for the low-energy part of the L$_3$ edge, as compared to experiment. First, the distance between the most intense feature in the low energy part of the spectrum to the {middle}-energy feature in the L$_2$ region is {7.2} eV, according to experiment, whereas 1e-X2C-B3LYP and DCB-X2C-EOM-CC predict significantly smaller values of {6.8} eV and {6.5} eV, respectively. {Both methods do a poor job of predicting the correct relative peak intensities. } The TDDFT-based analysis in Ref. \citenum{Prince09_2914} indicates that this feature around 521 eV and the other higher-energy features include both $L_2$ and $L_3$ edge features arising from 2${p_{1/2}}$ to valence and 2$_{p_{3/2}}$ to Rydberg orbital type excitations, respectively. {The lack of Rydberg basis function may be the source of such errors. }

\subsection{CrO$_{2}$Cl$_{2}$}
\label{SEC:Cr}

\begin{figure}[!htpb]
\includegraphics[width=\linewidth]{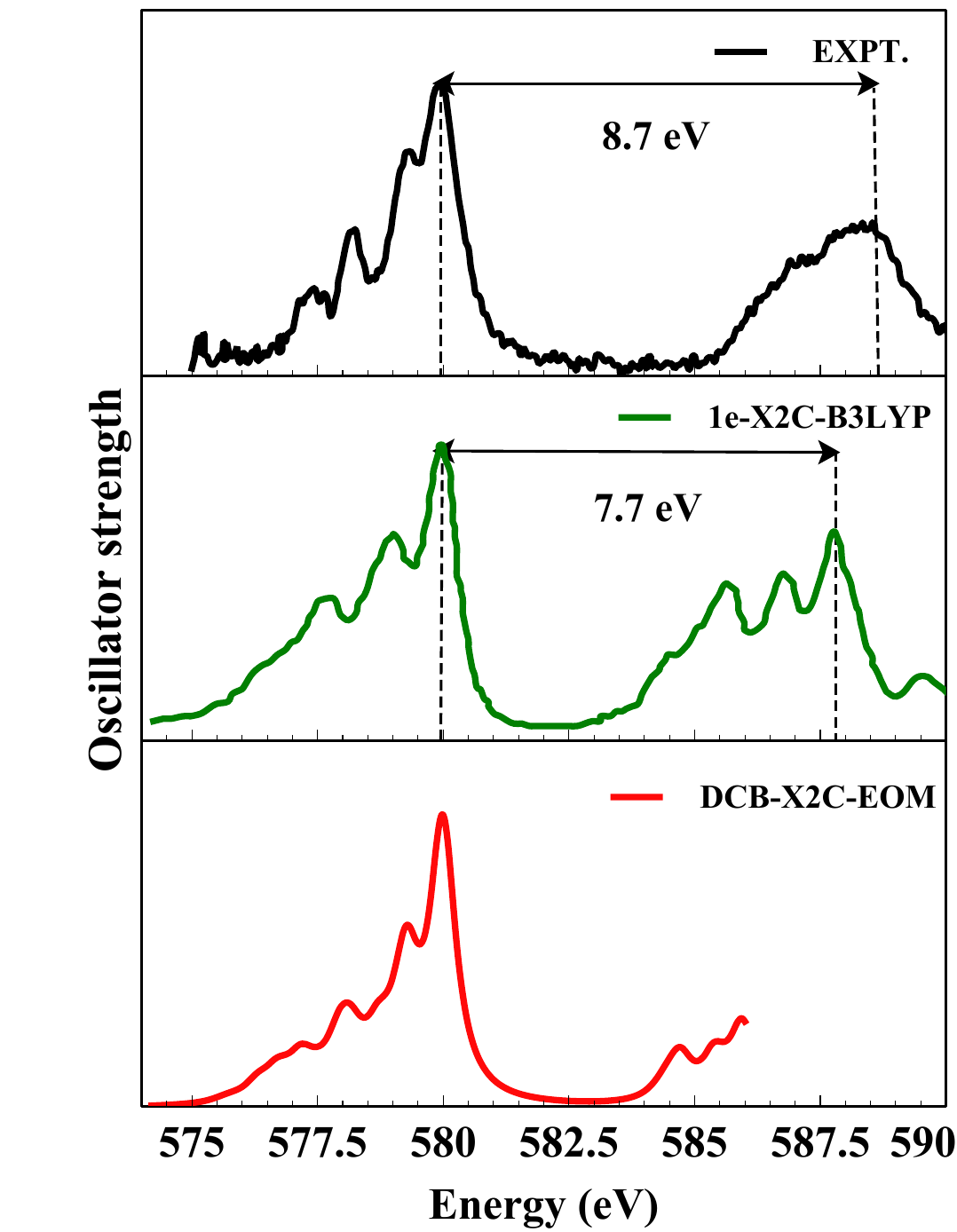}
 \caption{Comparison of L$_{2,3}$-edge spectra of CrO$_{2}$Cl$_{2}$ computed with 1e-X2C-TDDFT and DCB-X2C-EOM-CC using x2c-TZVPall-2c basis, alongside the experimental spectrum in the uppermost panel (reproduced from Ref.~\cite{Prince09_2914}). The 1e-X2C-TDDFT spectrum in the second panel obtained with the B3LYP functional and aug-cc-pVTZ basis has been reproduced from Ref. \citenum{Li18_1998}. The applied energy shifts are as follows: X2C-B3LYP: $+$8 eV, and DCB-X2C-EOM: $-${3.8} eV.
 }\label{fig:Cr_L}
\end{figure}

Lastly, we come to the case of CrO$_2$Cl$_2$, which, like  VOCl$_3$, has an L$_{2,3}$-edge absorption exhibiting two broad regions, arising from the combined effects of spin--orbit coupling and crystal field splitting. In this case, the reduction of symmetry from $T_d$ to $C_\text{2v}$ splits the Cr 3$d$ orbitals into five distinct virtual valence levels, resulting in an extremely high density of states spanning both the L$_3$ and L$_2$ edges.\cite{Prince09_2914} {These overlapping transitions result in an extremely high density of states, which leads to convergence difficulties and is why we fail to recover the highest-energy part of the spectrum at the CVS-DCB-X2C-EOM-CC level.} Again, 1e-X2C-B3LYP data can be obtained for the entire spectrum, but quantitative discrepancies between theory and experiment persist. L$_2$-edge relative peak intensities are reasonably described, but the splitting between the most intense L$_3$ and L$_2$ features is underestimated, relative to experiment by 1 eV (7.7 eV versus 8.7 eV). 

Despite the numerical challenges associated with capturing the high-energy portion of the L$_{2,3}$ edge, DCB-X2C-EOM-CC provides a description of the lower-energy portion that is of superior quality to that from 1e-X2C-B3LYP.  In \cref{fig:Cr_L}, the simulated spectra are shifted such that the most intense L$_3$ peak aligns with the experimental position at 580 eV. As has been observed in all other cases, 1e-X2C-B3LYP and DCB-X2C-EOM-CC require shifts of opposite signs; the shift required by 1e-X2C-B3LYP is $+$8 eV, while that for DCB-X2C-EOM-CC is significantly smaller in magnitude ({$-$3.8} eV). Close inspection of the L$_3$ region reveals that 1e-X2C-B3LYP has difficulty resolving four distinct features present in the experimental spectrum, whereas DCB-X2C-EOM-CC does an excellent job of recovering these features, in terms of both relative peak positions and intensities.

%% file: conclusions.tex
\label{SEC:CONCLUSIONS}

We have developed a relativistic core-valence-separated EOM-CC approach within the molecular mean-field X2C framework and applied the approach to the simulation of L-edge X-ray absorption spectra. By benchmarking against the Breit-Pauli-based perturbative approach,\cite{Coriani20_8314} we have confirmed the correctness of our implementation and also provided evidence of the robustness of the scheme to recontraction of the basis set after the mean-field step. Applications to first-row transition metal complexes has demonstrated that DCB-X2C-EOM-CC spetra improve over those from relativistic TDDFT, particularly in terms of global energy shifts and, in many cases, peak intensity ratios. As such, these results establish DCB-X2C-EOM-CC as a useful and broadly applicable tool for relativistic core-excitation spectroscopy.

We have also demonstrated some limitations of DCB-X2C-EOM-CC. First, for CrO$_2$Cl$_2$, the high density of states makes converging enough roots to completely span the L$_{2,3}$ region difficult. Second, the rather poor description of the separation between the L$_3$ and L$_2$ regions in VOCl$_3$, where DCB-X2C-EOM-CC underestimates the experimental separation by 1 eV, suggests that the CCSD model may not be sufficient for capturing this feature. Missing correlation effects from connected triple or higher-order excitations will likely improve this splitting, as well as provide a better description of the shake-up states.